\documentclass{IEEEtran}
\usepackage{cite}
\usepackage{amsmath,amssymb,amsfonts}
\usepackage{graphicx}
\usepackage{textcomp,nicefrac}
\usepackage{tabularx}
\usepackage[T1]{fontenc}
\usepackage{inconsolata}

\newcommand{\Italic}[1]{\textit{#1}}
\newcommand{\Codify}[1]{\texttt{\footnotesize{#1}}}
\newcommand{\Unitify}[1]{#1}
\newcommand{\Particle}[1]{#1}

\usepackage{listings}
\usepackage{xcolor}
\usepackage{here}

\definecolor{codegreen}{rgb}{0,0.6,0}
\definecolor{codegray}{rgb}{0.5,0.5,0.5}
\definecolor{codepurple}{rgb}{0.58,0,0.82}
\definecolor{backcolour}{rgb}{0.95,0.95,0.92}

\lstdefinestyle{mystyle}{
	backgroundcolor=\color{backcolour},   
	commentstyle=\color{codegreen},
	keywordstyle=\color{magenta},
	numberstyle=\tiny\color{codegray},
	stringstyle=\color{codepurple},
	basicstyle=\ttfamily\footnotesize,
	breakatwhitespace=false,         
	breaklines=true,                 
	captionpos=b,                    
	keepspaces=true,                 
	numbers=left,                    
	numbersep=5pt,                  
	showspaces=false,                
	showstringspaces=false,
	showtabs=false,                  
	tabsize=2
}

\graphicspath{ {./figures/} }

\def\BibTeX{
	{\rm B\kern-.05em{\sc i\kern-.025em b}\kern-.08em T\kern-.1667em\lower.7ex\hbox{E}\kern-.125emX}
}
\begin{document}
	
	\title{
		Remote Configuration of the ProASIC3 on the ALICE Inner Tracking System Readout Unit 
	}
	
	\author{
		Shiming Yuan on behalf of the ALICE Collaboration
		\thanks{
			Shiming Yuan is a Ph.D. candidate at the University of Bergen. (e-mail: shiming.yuan@uib.no).
		}
	}
	
	\maketitle
	
	\begin{abstract}
		A Large Ion Collider Experiment (ALICE) is one of the four major experiments conducted at the CERN Large Hadron Collider (LHC). 
		The ALICE detector is currently undergoing an upgrade for the upcoming Run 3 at the LHC. The new Inner Tracking System (ITS) sub-detector is part of this upgrade. 
		The front-end electronics of the ITS is composed by 192 Readout Units, installed in a radiation environment. 
		Single Event Upsets (SEUs) in the SRAM-based Xilinx Kintex Ultrascale FPGAs used in the ITS readout represent a real concern. 
		To clear SEUs affecting the Kintex configuration memory, a secondary Flash-based Microsemi ProASIC3E (PA3) FPGA is used. 
		This device configures and continuously scrubs the Xilinx FPGA while data-taking is ongoing, which avoids accumulation of SEUs. 
		The communication path to the RUs is via the radiation hard Gigabit Transceiver (GBT) system on 100 \Unitify{m} long optical links. 
		The PA3 is reachable via the GBT Slow Control Adapter (GBT-SCA) ASIC using a dedicated JTAG bus driving channel.
		
		During the course of Run 3, it is foreseeable that the FPGA design of the PA3 will require upgrades to correct possible issues and add new functionality. 
		It is therefore mandatory that the PA3 itself can be configured remotely, for which a dedicated software tool is needed. This paper presents the design and implementation of the distributed tools to re-configure remotely the PA3 FPGAs.
	\end{abstract}
	
	
	\section{Introduction}
	
	\label{sec:introduction}
	
	\IEEEPARstart{T}he ALICE experiment is designed to study the strong interactions between \Particle{quarks}, in particular the properties of Quark-Gluon Plasma (QGP), through measurements and observations of heavy ion collisions at the LHC.
	The LHC is undergoing an upgrade during Long Shutdown 2 (LS2), which started in December 2018. 
	This upgrade will bring the target integrated luminosity of \Particle{Pb-Pb} collisions to 10 \Unitify{nb$^{-1}$} at the design LHC energy of $\sqrt{s_{NN}}$=5.5\Unitify{TeV}.
	This higher luminosity represents an increase of a factor of 10 in the inspected luminosity for rare signals, and a factor of 100 for the minimum-biased data sample\cite{abelev2014technical}. 
	The major upgrade that the ALICE apparatus is undergoing has therefore to cope also with the increased data rate. 
	During LS2, a completely new detector will replace the Inner Tracking System (ITS), with significantly reduced material budget while delivering higher detector resolution.
	It is based on the ALPIDE (\textbf{AL}ice \textbf{PI}xel \textbf{DE}tector, a custom-made CMOS monolithic active pixel sensor) chip sensors, which are controlled and read-out by the Readout Units (RUs) in ITS.
	\subsection{The upgraded ITS}
	
	\begin{figure}[H]
		\centering
		\includegraphics[width=3.5in]{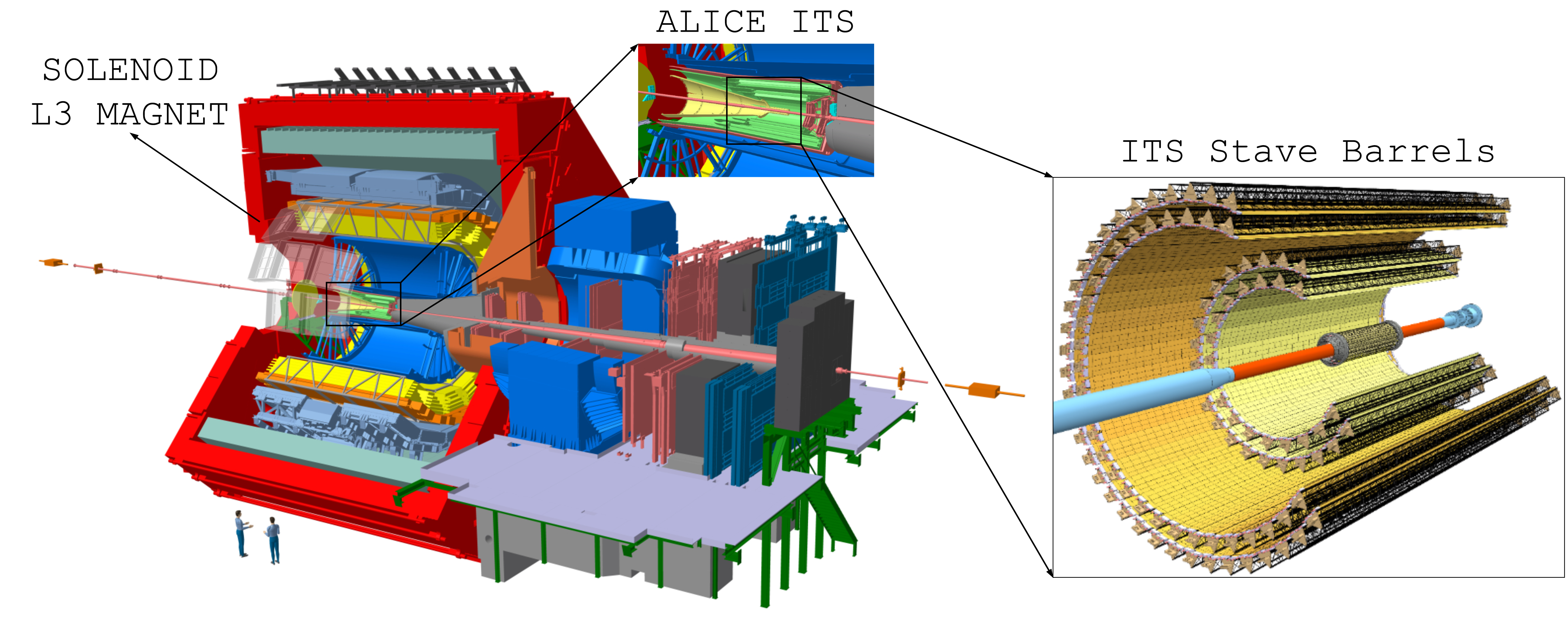}
		\caption{ALICE and The ALICE Inner Tracking System}
		\label{fig:ALICE and ITS}
	\end{figure}	
	
	The upgraded ITS consists of seven concentric cylindrical layers of ALPIDE arrays surrounding the beam pipe at distances from the Interaction Point (IP) ranging from 22 \Unitify{mm} at the inner most layer to 430 \Unitify{mm} at the outer most layer. 
	The location of the Inner Tracking System in ALICE is shown in Figure \ref{fig:ALICE and ITS}.
	
\begin{figure}[H]
	\centering
	\includegraphics[width=3.5in]{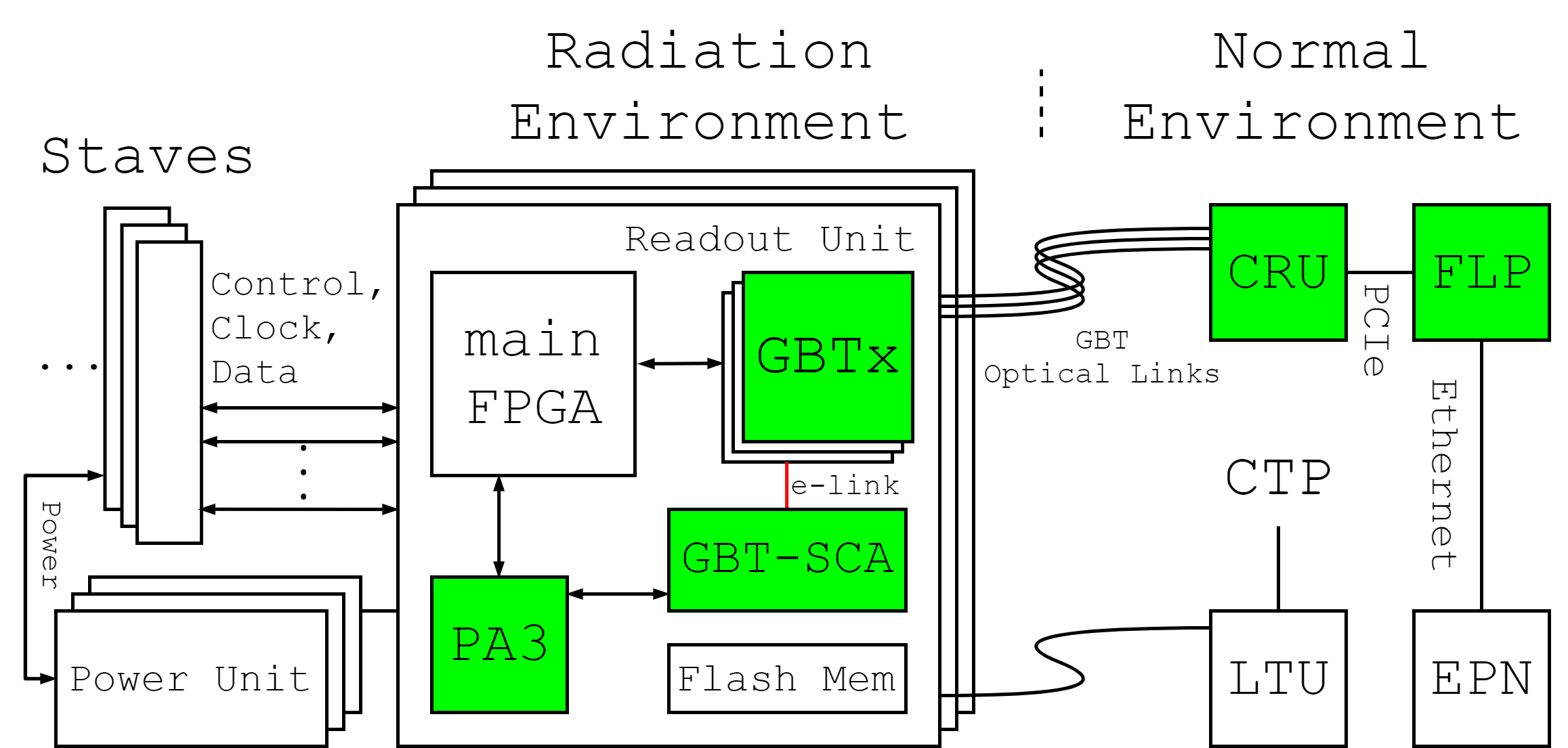}
	\caption{The ALICE ITS Readout System}
	\label{fig:ALICE ITS Readout System}
\end{figure}
	
	Figure \ref{fig:ALICE ITS Readout System} shows the block diagram of how the RU links the front-end sensor with the other parts of the ITS readout chain. The RUs are located in a radiation environment.
	The RU ships the data to the Common Readout Unit (CRU)  installed in the counting room, in a radiation safe environment.
	The Xilinx Kintex UltraScale FPGA (main FPGA) on the RU is responsible for the data streaming, trigger handling, and control of the sensors. 
	It also interfaces a dedicated power board which controls the detector powering.
	To avoid compromising the data taking and detector safety it is therefore critical to avoid radiation induced functional failures on the RU main FPGA.
	
	\begin{figure}[H]
		\centering
		\includegraphics[width=3.5in]{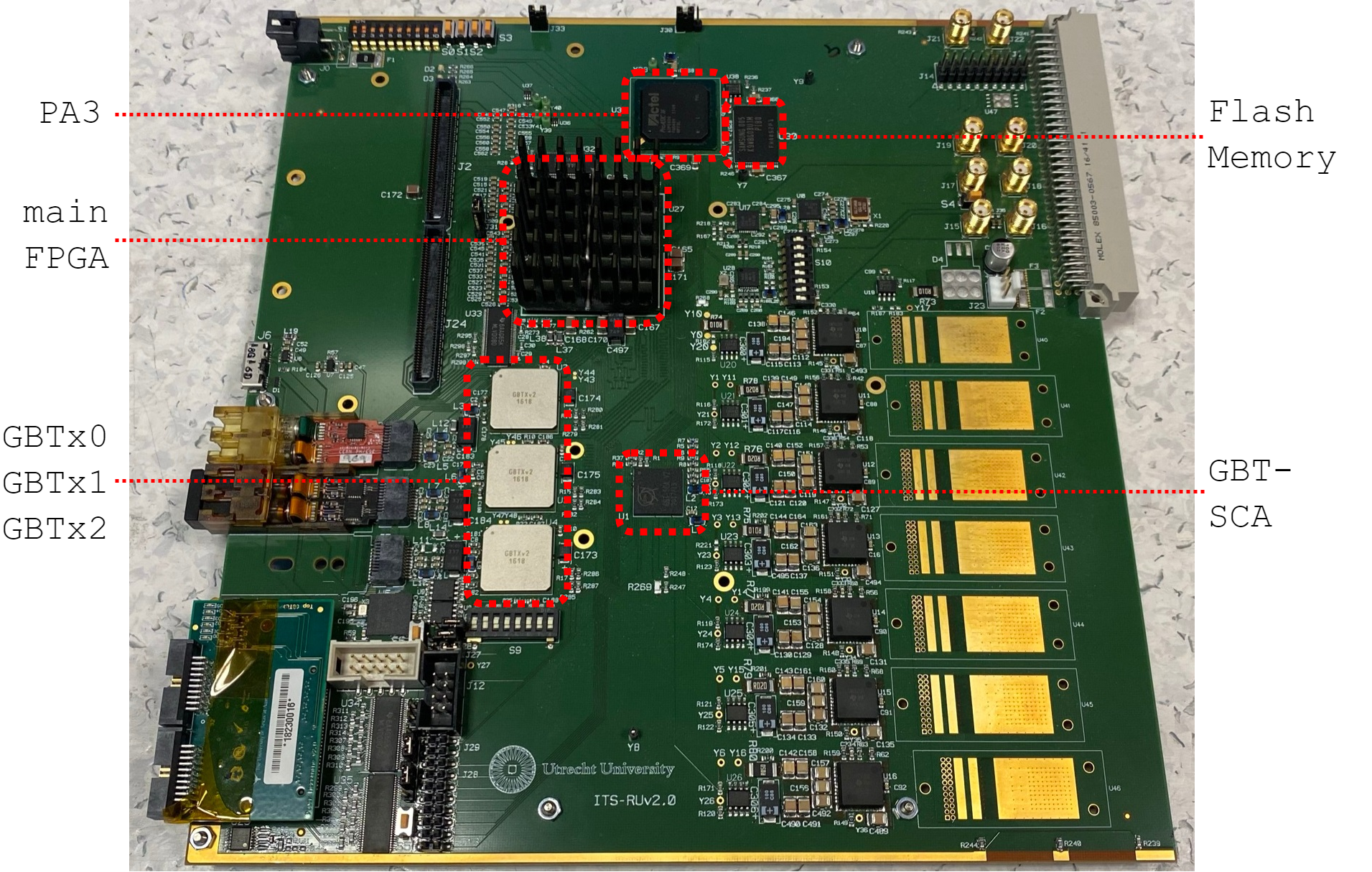}
		\caption{The ALICE ITS Readout Unit}
		\label{fig:ALICE ITS Readout Unit}
	\end{figure}

	SEU-mitigation techniques such as Triple Modular Redundancy (TMR), which is implemented in the main FPGA, and “scrubbing”, which is featured by the secondary Flash-based Microsemi ProASIC3 (PA3) FPGA, are adopted to minimize the impact of the radiation-induced SEUs in the main FPGA. 
	The anticipated high-energy hadron flux in the environment where the RUs are installed is around 1 \Unitify{kHz/cm$^2$}, and the PA3 is capable of reconfiguring the main FPGA in a scrub cycle of 1.7\Unitify{s}\cite{ersdal2019external}. 
	The cooperation of these SEU-mitigation techniques will safe guard the RUs against the radiation-induced functional failures during the run.
	The RUv2.0 used in this work is shown in Figure \ref{fig:ALICE ITS Readout Unit}, and the main components on the board with respect to this paper are shown in the dotted boxes, on top of the main FPGA is the cooling device.
	
	The radiation-hardened Gigabit Transceiver (GBT) optical links that connect RUs and CRUs are capable of transmitting one 120-\Unitify{bit} GBT frame per LHC bunch crossing interval ($\approx$25\Unitify{ns}), resulting in a line data rate of 4.8 \Unitify{Gbps}. 
	Two additional bits in the GBT frame are the External Control (EC) field\cite{moreira2018gbtxmanual}.
	The EC is passed to the GBT-SCA from the GBT interface. 
	This field carries the configuration data for PA3.
	
	The CRU sits in the First Level Processors (FLPs) in the counting room.
	The FLP hosts software for control and data read-out, and is responsible of forwarding the aggregated data from different links over Ethernet to the Event Processing Nodes (EPNs) which stores the data for further analysis.
	Each FLP is capable of hosting up to two CRUs on its PCI Express slots, and each CRU is capable of linking up to eight RUs.
	
	The CRU is responsible for data aggregation and control of the front-end electronics. 
	The CRU memory space is PCIe BAR-mapped by the Portable Driver Architecture (PDA) driver via the ReadoutCard (RoC) module of the ALICE Online-Offline computing system (ALICE O$^2$).
	The RoC module is a C++ library that provides a high-level interface for accessing and controlling data acquisition PCIe cards\cite{aliceo2readoutcard}. It offers an API to the PCIe BAR.
	
	\subsection{Motivation for remote configuration of the PA3}
	
	The remote configuration application for the PA3, \Codify{pa3jtag}, is a dedicated software tool that is foreseen to be used by detector experts only in periods when data taking is not ongoing. 
	It uses the same hardware and driver structures as the central Detector Control System (DCS), therefore a low level arbitration is needed. 
	This exists in the ALICE O$^2$\cite{aliceo2LLA}, and the \Codify{pa3jtag} application utilizes this to ensure safe detector operation. 
	
	It is not foreseen that remote configuration of the PA3 will be a frequent operation.
	It is however important to have the opportunity to remotely upgrade the PA3 for potential future upgrades or bug-fixes. 
	Special PA3 builds also exist that enable for instance fault injection, which can be used to qualify the robustness of new versions of the main FPGA firmware. 
	Additionally, the physical access to the RUs are limited due the radiation environment and the actual location of the installation, so it is of vital importance to have a remote configuration functionality.
	
	\subsection{Structure of the paper}
	This paper presents \Codify{pa3jtag}, a new software tool that is used to reconfigure the PA3 FPGA on the ALICE ITS Readout Unit for the ITS experts.
	The structure of the paper is:
	\begin{enumerate}
		\item Section I gives an overview the of upgraded ALICE Inner Tracking System, and the motivation for developing \Codify{pa3jtag}.
		\item Section II describes the hardware components related to \Codify{pa3jtag} in ITS.
		\item Section III introduces \Codify{DirectC}, the software tool based on which \Codify{pa3jtag} has been developed.
		\item Section IV lays out the design and implementation of \Codify{pa3jtag}.
		\item Section V summarizes the work.
	\end{enumerate}

	\begin{figure*}
		\centering
		\includegraphics[width=\textwidth]{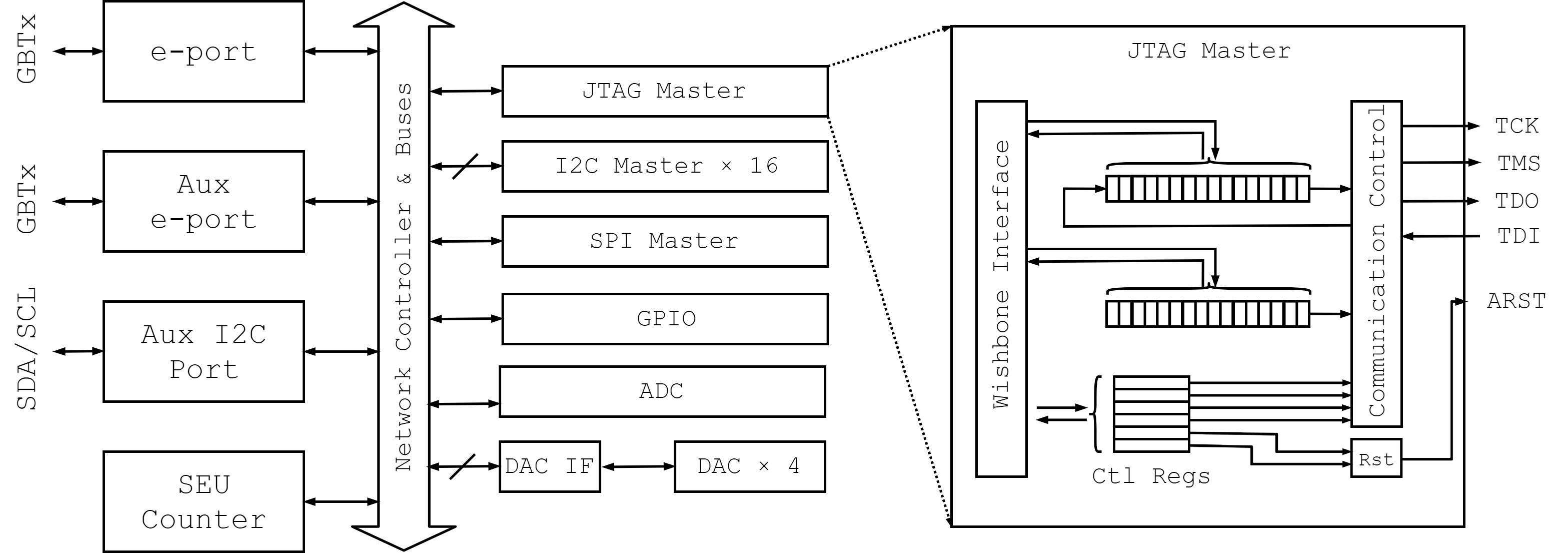}
		\caption{GBT-SCA Block Diagram, and the JTAG master module}
		\label{fig:GBT-SCA Block Diagram}
	\end{figure*}
	
	\section{System Components and Datapath of the PA3 Configuration}
	
	
	\subsection{The GigaBit Transceiver - Slow Control Adapter}
	The PA3 configuration data is transmitted via the 2-bit EC field in the GBT Frame from the CRU to the RU.
	As shown in Figure \ref{fig:ALICE ITS Readout System}, the GBT-SCA is connected to a dedicated port on GBTx0 through e-links, which is a 80 \Unitify{Mbps} dual redundant bidirectional data-link. 
	GBTx0 passes the EC information to GBT-SCA via the e-links.
	The GBT-SCA has several user-configurable interface ports: 1 SPI master, 16 independent 1$^2$C masters, 1 JTAG master, and 32 General Purpose IOs, as is shown in the GBT-SCA block diagram in Figure \ref{fig:GBT-SCA Block Diagram}.
	The GBT-SCA move the EC data from the E-PORT to the Network Controller in a parallel bus, and the Network Controller further distributes the data to the specified interface channel via a Wishbone bus.
	The GBTx is transparent to the communication between the CRU and the GBT-SCA. The GBT-SCA interface channels are memory mapped on the PCIe bus, thus accessible from the RoC API. The JTAG Master channel is connected to the RU JTAG configuration chain, hence it is utilized for the configuration of PA3.
	
	The JTAG channel configuration registers in the GBT-SCA are described in Table \ref{table:JTAG Channel Configuration Registers}. 
	The JTAG channel of GBT-SCA has two 128-bit shift registers that serializes and deserializes the data between the JTAG lines and the internal modules.
	The TDO and TDI shift registers physically share the same hardware registers, though the assigned addresses in the PCIe BAR space differ.
	The control register defines the number of bits transmitted in a single JTAG operation, the MSB/LSB, and serves as the busy flag of the JTAG bus.
	The vacancy of a JTAG TAP controller in the GBT-SCA means the JTAG data packets will be generated by an agent deployed on the other end of the GBT optical links.
	
	A typical JTAG command that writes to the TDI line consists of the following steps:
	
		\begin{enumerate}
			\item Setting the TDI data packet length by writing to the control register
			\item Writing the TDI data to the TDI shift register, if the data is longer than 32 bits, then multiple writes to adjacent sections of the buffer are necessary
			\item Write to the JTAG$\textunderscore$GO register, which starts the serial communication on the JTAG bus
		\end{enumerate}
	
	\begin{table}[hbt!]
		\centering	
		\caption{JTAG Channel Configuration Registers}
		\setlength{\tabcolsep}{5pt}
		\begin{tabular}{|m{1.2cm}|m{0.6cm}|m{3.5cm}|m{1.1cm}|}
			\hline
			Register & Mode & Function & Size (bit)\\
			\hline
			Control  & R/W  & Define operation mode of the channel & 16\\ 
			\hline
			Frequency & R/W & Set the operating frequency of the channel & 16\\
			\hline
			TDO  & R/W       & Data transmit buffer for the TDO line & 32$\times$4\\
			\hline
			TDI  & R        & Data receive buffer for the TDI line & 32$\times$4\\
			\hline
			TMS  & R/W        & Data transmit buffer for the TMS line & 32$\times$4\\
			\hline
		\end{tabular}
		\label{table:JTAG Channel Configuration Registers}
	\end{table}		
	
	\subsection{The RU JTAG Configuration Scheme}
	
\begin{figure}[hbt!]
		\centering
		\includegraphics[width=3.5in]{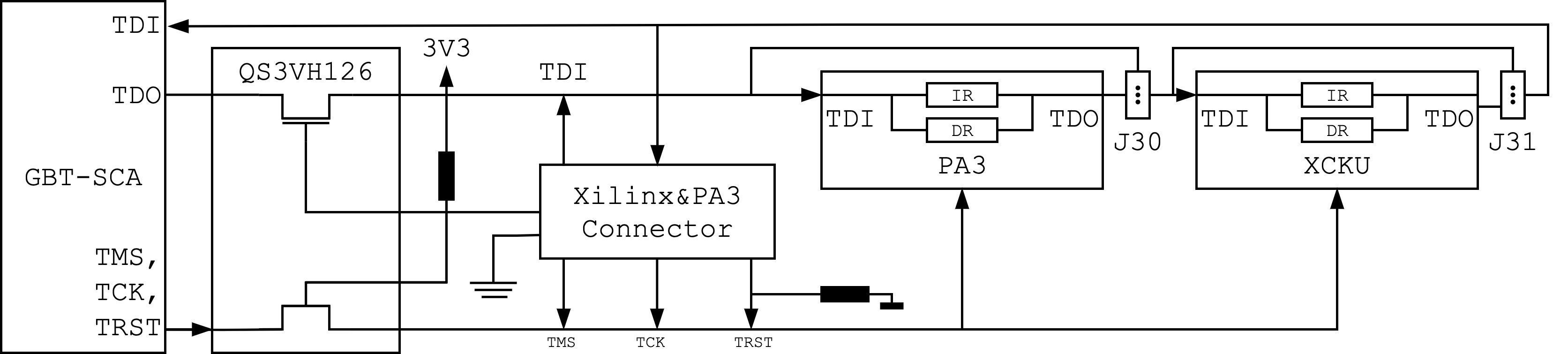}
		\caption{Readout Unit JTAG Configuration Scheme}
		\label{Fig:Readout Unit JTAG Configuration Scheme}
	\end{figure}

	The JTAG configuration datapath on the RU is shown in Figure \ref{Fig:Readout Unit JTAG Configuration Scheme}. 
	The TMS pins of PA3 and the main FPGA are connected in parallel with the TMS output pin of GBT-SCA; whereas the TDI/TDO line is connected in serial. 
	The Instruction Registers (IR) and Data Register (DR) are connected in a long shift register from the TDO to the TDI pins of GBT-SCA.
	The IR lengths for PA3 and the main FPGA are 4, and 6 respectively.
	The DR length depends on the instruction in the IR, and when the device is in bypass mode, the DR length is set to 1.
	
	\section{\Codify{DirectC}}
	
	\Codify{DirectC} is a suite of C/C++ code designed for the In-System Programming (ISP) of several series of Microsemi FPGA products\cite{directcuserguide}. 
	Users need to make modifications to the software suite taking different system constraints into account.
	A typical ISP setup consists of three major parts: a JTAG Test Access Port (TAP) controller, a memory space storing the configuration file, and the JTAG configuration chain where the programmable devices are connected end-to-end. 
	\Codify{DirectC} serves as the JTAG TAP controller in a way that it programs the processor to toggle certain spaces in the memory that is designated to the output port, for instance the GPIO pins, which is physically connected to the JTAG pins on the JTAG configuration chain in system. \Codify{DirectC} by default supports bit-banging of the JTAG signal protocol, which is inefficient in the context of Slow Control communication, so it must be adapted to use the buffer register in the JTAG master module of GBT-SCA.
	
	\subsection{The \Codify{.dat} Configuration File}
	
	The \Codify{.dat} configuration file is generated by the Microsemi Libero development tool, and is preferred by the \Codify{DirectC} toolset. The \Codify{.dat} file for PA3 contains the following sections: 
	
	\begin{enumerate}
		\item Header Block - Identifies the type of the image, the size of the image, device information, and different flags that guide the \Codify{DirectC} algorithm to identify which block is supported and its associated options.
		\item Data Lookup Table - Pointers to the starting of the relative location of all the different data blocks used in \Codify{DirectC} algorithm, and the size of each block.
		\item Data Block - The configuration binary data for the data blocks pointed in the Data Lookup Table.
	\end{enumerate}
	
	The sizes, and the main information of each section are given in Table \ref{Table:Format Description of PA3 Configuration File}\cite{directcuserguide}.
	
	\begin{table}[hbt!]
		\centering
		\caption{Format Description of PA3 \Codify{.dat} File }
		\label{table}
		\setlength{\tabcolsep}{5pt}
		\begin{tabular}{|m{0.5cm}|m{1.4cm}|m{5cm}|}
			\hline
			& Size (Byte) & Information \\
			\hline
			HB  & 69           & Designer version number, Header size, Image size, Compression Flag, etc.. \\ 
			\hline
			DLT & n$_1$$\times$ 9 & Data Identifier 1, Pointer to data 1 memory location in the data block section, \# of bytes of data 1; Data Identifier 2, Pointer to data 2 memory location in the data block section, etc.. \\
			\hline
			DB  & n$_2$ + 2        & Configuration binary data, CRC of the entire image \\
			\hline
		\end{tabular}
		\label{Table:Format Description of PA3 Configuration File}
	\end{table}
	
	The size of DLT is n{$_1$} $\times$ 9 \Unitify{Bytes}, where n{$_1$} is the number of data pointers; the size of DB is n{$_2$} + 2 \Unitify{Bytes}, where n{$_2$} is the size of the configuration binary data, and the final 2 \Unitify{Bytes} are CRC of the entire image.
	Over 99.9$\%$ of the \Codify{.dat} file is the configuration binary data.
	
	\subsection{The Algorithm of \Codify{DirectC}}
	
	The entry function of \Codify{DirectC} is \Codify{dp$\_$top}.
	\Codify{dp$\_$top} passes an \Codify{Action$\_$code} to \Codify{dp{$\_$}top$\_$g3}. 
	\Codify{dp{$\_$}top$\_$g3} initializes the system for configuration, and invokes \Codify{dp$\_$perform$\_$action} to execute the corresponding algorithms based on the given \Codify{Action$\_$code} in a switch statement.
	
	For example, when the program tries to read the device id information, \Codify{dp$\_$perform$\_$action} would invoke \Codify{dp$\_$read$\_$idcode}, and the pseudo code for this function is shown below:
	
	\begin{lstlisting}[language=C++]
		void dp_read_idcode(void)
		{
			opcode = IDCODE;
			IRSCAN_in();
			goto_jtag_state(JTAG_RUN_TEST_IDLE);
			DRSCAN_out(IDCODE_LENGTH, global_buf);
			device_ID = mem_shift(global_buf);
			return;
		}\end{lstlisting}
	
	The \Codify{dp$\_$read$\_$idcode$\_$action} function consists of three consecutive software-hardware interactive steps:
	
	\begin{enumerate}
		\item The global variable \Codify{opcode} is assigned with the \Codify{IDCODE}, and \Codify{IRSCAN$\_$in} (IR stands for Instruction Registers) shifts the \Codify{opcode} to the TDO/TDI line of the JTAG chain;
		\item \Codify{goto$\_$jtag$\_$state} shifts the corresponding TMS packet to invoke JTAG state changes;
		\item \Codify{DRSCAN$\_$out} (DR stands for Data Registers) shifts the TDO read-back TDO/TDI line to the memory space controlled by the processor.
	\end{enumerate}
	
	The \Codify{Action$\_$code} look-up table is pre-defined by the vendor. To verify the \Codify{device$\_$ID} read-back, the \Codify{DirectC} asserts it with the corresponding information stored in the \Codify{.dat} file. The relative location of this information is pre-defined in \Codify{DirectC} header files.
	
	The functions: IR scan-in, DR scan-out, and the JTAG state-change all invoke \Codify{dp$\_$jtag$\_$tms$\_$tdi$\_$tdo}, which calculates the JTAG pin digits, assigns them to a global 8-bit static variable \Codify{char jtag$\_$port$\_$reg}, and passes it to \Codify{jtag$\_$outp}. \Codify{jtag$\_$outp} sets the JTAG port by toggling the specified memory space, thus putting the configuration digits on the JTAG lines.
	
	\section{\Codify{pa3jtag} Design and Implementation}
	
	The goal of \Codify{pa3jtag} is to configure the PA3 via the Slow Control datapath by utilizing the PCIe BAR-mapped JTAG channel of GBT-SCA from the FLP. 
	Simply put, configuration of an FPGA consists in writing the configuration file into the FPGA configuration memory, after having passed through the preparation steps such as setting the JTAG transmitting frequency, verifying the JTAG chain, checking the device ID \Italic{etc.}.
	
	Initially, various options for remote configuration of the PA3 were investigated, such as writing the software from scratch, or adapting the \Codify{STAPL Player} code.
	
	Making the tool from scratch was infeasible for two main reasons: 
	(1) Configuring the PA3 requires taking various parameters of the device defined by the vendor into the program, and these parameters are usually defined in the compatible configuration tools maintained by the vendor. So the development was to rely partly on the open-source code released by the vendor, and (2) the schedule of the project forbade the developing-from-scratch approach.
	
	The Standard Test and Programming Language (STAPL) was first introduced in 1999 to support the programming of programmable devices and testing of electronic systems, using the IEEE Standard 1149.1: "Standard Test Access Port and Boundary Scan Architecture" (commonly referred to as JTAG). As a STAPL file is executed, signals are produced on the IEEE 1149.1 interface, as described in the STAPL file\cite{stapl1999}.
	The \Codify{STAPL Player} is a interpreter program that executes the statements in the STAPL file directly without compiling these statements into a binary executable file\cite{stapl1999}.
	
	The earliest \Codify{STAPL Player} available was released by Actel in 2003 with Application Note AC171\cite{actelstapl2003}.
	And the most recent Actel \Codify{STAPL Player} (v1.1) was released in 2008\cite{actelstapl2008}\footnote{Actel was acquired by Microsemi in 2010.}. 
	Though it was tried to modify \Codify{STAPL Player} v1.1 for configuring PA3, and performing few other actions such as scanning the JTAG chain, reading the device information worked with the modified code, but the programming action failed with syntax errors in the STAPL file that was generated by the Microsemi Libero tool.
	
	The LHCb team who worked on a similar agenda encountered similar issues with the Actel \Codify{STAPL Player}, and diverted to \Codify{DirectC}\cite{lhcb}.
	Other \Codify{STAPL Player} options, such as the Intel \Codify{Jam STAPL Player}, are not compatible with the PA3 STAPL files generated by Microsemi Libero.
	
	It turned out that basing the \Codify{pa3jtag} on \Codify{DirectC} v4.1 was the best option for remote configuration. 
	As introduced in section III, \Codify{DirectC} is a tool that is capable of parsing and interpreting a \Codify{.dat} file to JTAG Test Mode Select (TMS) and Test Data In (TDI) bitstreams, and writing/reading the TMS, TDI/TDO (Test Data Out) sequential bits to/from the JTAG ports\cite{directcuserguide}. 
	As mentioned in section II-B, the JTAG chain on the RU has two components: the main FPGA and the PA3. 
	The configuration file in \Codify{DirectC} sets the data register and instruction register lengths of both devices. 
	This ensures that the configuration file is written correctly to the target FPGA.
	
	What \Codify{DirectC} lacks is the utilization of the upper part of the configuration datapath from the FLP to GBT-SCA.
	Therefore a \Codify{GBT$\textunderscore$SCA} class that utilizes the GBT-SCA JTAG channel has been designed based on the GBT-SCA user manual\cite{gbtscamanual}. 
	
	\textbf{The GBT-SCA interface \Codify{class GBT$\_$SCA}}\indent 
	As introduced in section II-A, the JTAG channel of GBT-SCA includes two 128-\Unitify{bit} shift registers as transmit buffers that serializes and deserializes the configuration TMS and TDI/TDO packets.
	In order to accelerate the configuration of the PA3, multiple modifications have been applied to the \Codify{DirectC} algorithm so that it uses the transmit buffers in the JTAG channel. 
	The 128-\Unitify{bit} transmit buffers, each of which consists of four 32-\Unitify{bit} registers, and the CONTROL and FREQUENCY registers are accessible respectively from \Codify{GBT$\textunderscore$SCA} class.
	The \Codify{GBT$\textunderscore$SCA} class implements the option of setting the optimal size of the packets prior to writing it on the JTAG line. 
	The packets have different sizes in different parts of the JTAG bitstream, and some requires verifying the TDO read-backs, while others do not. 
	The pseudo code for sending a data packet to the JTAG line is shown below:
	
	\begin{lstlisting}[language=C++]
		void jtag_tr(u128_t tms, u128_t tdi, u32_t length)
		{
			if(length != G_length)
			{
				G_length = length;
				jtag_w_CTRL(G_length);
			}
			if(tms != G_tms) 
			{
				G_tms = tms_data;
				jtag_w_TMS(G_tms);
			}
			jtag_w_TDI(tdi);
			jtag_start_transmission();
		}\end{lstlisting}
	
	In order to minimize the number of accesses to the JTAG channel, some of the data to be transmitted are compared with global variables storing the data from the previous transmission. 
	If the data to be transmitted is the same as the last transmission, then \Codify{pa3jtag} omits this transmission. The TDO/TDI register is always used to transmit different data, so this mechanism is only implemented for TMS buffer and the control register. 
	
	Apart from accessing the transmission and control registers, the GBT-SCA interface also provides functions on an operational level, such as power switch of the JTAG channel, setting JTAG transmitting frequency, GBT-SCA id verification, and GBT-SCA initialization, etc.
	
	\textbf{Concatenating data packets in JTAG state shifts}\indent 
	The function \Codify{IRSCAN$\textunderscore$in}, which sends an \Codify{opcode} to the JTAG lines, is a frequently used function in the \Codify{DirectC} algorithm. 
	It consists of three TMS writes and one TDI write. 
	Sending each of the TMS/TDI packet requires one write to the GBT-SCA JTAG transmit buffer and one write to the \Codify{JTAG$\textunderscore$GO} command register, therefore eight GBT-SCA accesses were required in the original algorithm. 
	Depending on different \Codify{opcode}s, the TDI packets vary, but the rest stays the same. 
	Plus, this function does not require TDO read-backs. 
	Based on this observation, the algorithm has been modified into a way that the packets generated in \Codify{IRSCAN$\textunderscore$in} are concatenated and written to the JTAG transmit buffer with three GBT-SCA writes (one TDI packet, one TMS packet, and one \Codify{JTAG$\textunderscore$GO} command), thus reducing the number of accesses to GBT-SCA as optimally few as possible. 
	Often, the length of the packet involved in \Codify{IRSCAN$\textunderscore$in} is less than 32-\Unitify{bit}, in which case it is sufficient in these cases to utilize only the least-significant section of the transmit buffer.
	This strategy has also been applied to modify other functions in the \Codify{DirectC} algorithm in the similar manner.
	
	\textbf{Accelerating \Codify{program$\textunderscore$array}}\indent The most time-consuming part of the PA3 configuration is writing the configuration data on the Look-Up Table (LUT) array. 
	In the PA3, the 3444-Row (also know as Column Grid, or CG) memory block array stores the configuration. 
	Based on our observation of the JTAG bitstream involved with \Codify{program$\textunderscore$array}, the total length of the TDI packets corresponding to one CG is 98-\Unitify{bit}, therefore the algorithm has been modified in a way to concatenate all the involved packets to a single 98-\Unitify{bit} packet, and utilizes all the four section of the 128-\Unitify{bit} shift register for the transmission. 
	As for the TDO read backs involved with \Codify{program$\textunderscore$array}, tests and observations did show that the TDO read-backs are constants representing device information and configuration status, instead of variables depending on the configuration data. 
	Therefore the algorithm has been modified for \Codify{pa3jtag} to skip the execution of TDO read-backs, and provided with ``fake TDO read-back'' constants to further accelerate the configuration process.
		
	\section{Comparison  and Summary}

A similar problem of controlling the Front-end Electronics (FE) from the Experiment Control System (ECS) via the GBT-SCA chip was encountered by one of the sub-detector teams at the LHCb experiment. 
They have a different approach. 
In the LHCb solution, a dedicated interface board (SOL40) is designed to distribute all the control signals to the FEs\cite{sol402018barbosa}.
In the SOL40's firmware, the SOL40-SCA core is responsible of sending ECS commands to the GBT-SCA.

\begin{figure}[H]
	\centering
	\includegraphics[width=3.5in]{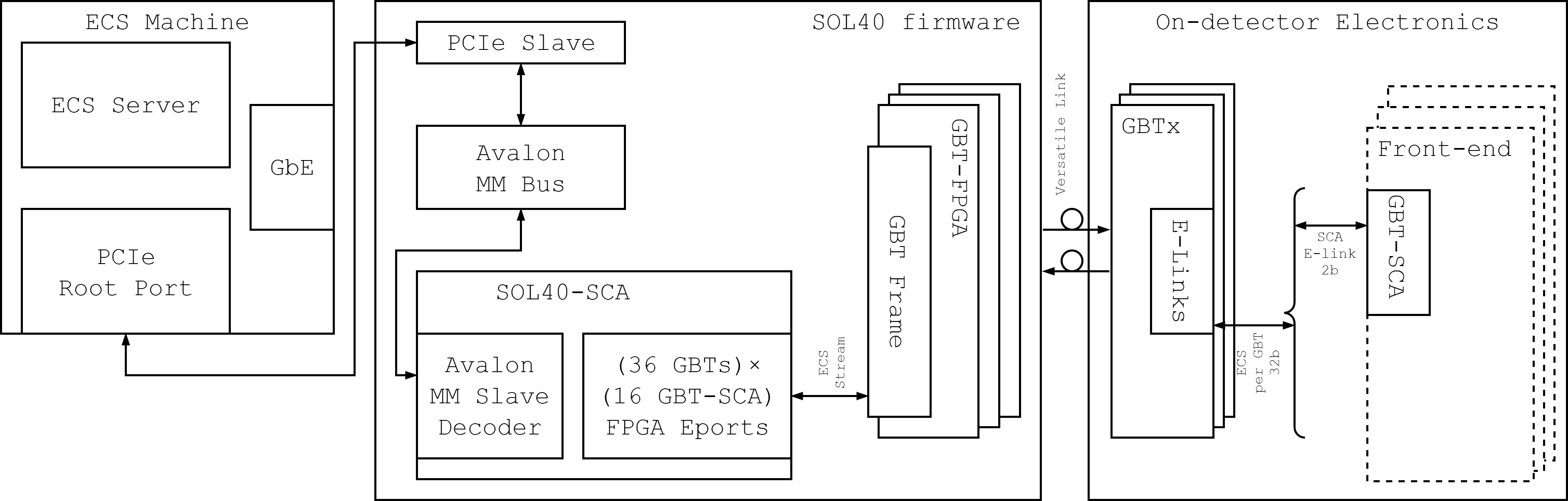}
	\caption{The LHCb Solution}
	\label{Fig:The LHCb Solution}
\end{figure}

	Figure \ref{Fig:The LHCb Solution} shows the block diagram of how the SOL40-SCA core interfaces between the LHCb ECS and FEs.

	When configuring the FPGAs on the FE, the SOL40-SCA core serves as a JTAG TAP controller, thus it significantly lowers the time consumption for the FPGA JTAG configuration action, compared to our solution. 

	The ALICE CRU is at the same hierarchical position as SOL40 in LHCb, however an interface like the SOL40-SCA is not added in the CRU firmware, hence the pa3jtag needed to be implemented as a pure software tool. 
	If a similar functionality as the SOL40-SCA was added to the CRU firmware, the ITS team would be responsible for the maintenance of this ITS specific module, which opposes to the current situation where the CRU team is fully in charge of the CRU firmware.
	The complexity of the CRU design would increase if sub-detector specific modules are added. The manpower to develop and provide maintenance to this module would also be something to consider. 
	Therefore a pure software solution is reasonable for our application.
	
	In conclusion, the size of the \Codify{.dat} file for PA3 is 537,694 \Unitify{Bytes}, and the bottleneck of the configuration is the access to the GBT-SCA JTAG channel registers from the FLP, which takes on average 3 \Unitify{ms}. Although various techniques have been applied to accelerate the configuration, the unavoidable TDO read-backs and the transmission of smaller packets significantly slow down the process. 
	As of now, \Codify{pa3jtag} is able to configure the PA3 in 31 minutes, implying that the effective configuration bit rate is around 2.4 \Unitify{Kbps}. 
	The time consumption is however acceptable for the contextual usage.
	In principle, \Codify{pa3jtag} can also be used to configure the main FPGA. 
	But the size of the configuration file for the main FPGA is 24,125,021 \Unitify{Bytes}, $\approx$45 times to the PA3.
	The configuration time of the Xilinx using the PA3 and the flash memory is approximately 2 s, therefore it makes no sense to use the Slow Control data path for the configuration of the main FPGA. 
	
	However, since the PA3 configuration action is not considered to be frequently executed, and it is possible to run it in parallel on each individual FLP, it is considered to be fast enough in our case. 
	The possibility to configure the PA3 on the RU is anyway considered an important component to ensure the robustness of the real-time data taking system the ITS RU represents.
	
	\begin{flushleft}
		\bibliographystyle{IEEEtran}
		\bibliography{bibliography}

\begin{thebibliography}{10}
\providecommand{\url}[1]{#1}
\csname url@samestyle\endcsname
\providecommand{\newblock}{\relax}
\providecommand{\bibinfo}[2]{#2}
\providecommand{\BIBentrySTDinterwordspacing}{\spaceskip=0pt\relax}
\providecommand{\BIBentryALTinterwordstretchfactor}{4}
\providecommand{\BIBentryALTinterwordspacing}{\spaceskip=\fontdimen2\font plus
\BIBentryALTinterwordstretchfactor\fontdimen3\font minus
  \fontdimen4\font\relax}
\providecommand{\BIBforeignlanguage}[2]{{%
\expandafter\ifx\csname l@#1\endcsname\relax
\typeout{** WARNING: IEEEtran.bst: No hyphenation pattern has been}%
\typeout{** loaded for the language `#1'. Using the pattern for}%
\typeout{** the default language instead.}%
\else
\language=\csname l@#1\endcsname
\fi
#2}}
\providecommand{\BIBdecl}{\relax}
\BIBdecl

\bibitem{abelev2014technical}
{B. Abelev et al. (ALICE Collaboration)}, ``{Technical design report for the
  upgrade of the ALICE inner tracking system},'' \emph{Journal of Physics G:
  Nuclear and Particle Physics}, vol.~41, no.~8, p. 087002, 2014.

\bibitem{ersdal2019external}
M.~R. Ersdal, J.~Alme, M.~Bonora, P.~Giubilato, M.~Lupi, S.~V. Nesb{\o}, A.~U.
  Rehman, D.~R{\"o}hrich, G.~A. Rinella, J.~Schambach, A.~Velure, and S.~Yuan,
  ``External scrubber implementation for the alice its readout unit,'' in
  \emph{Topical Workshop on Electronics for Particle Physics TWEPP2019},
  vol.~2, 2019, p.~6.

\bibitem{moreira2018gbtxmanual}
P.~Moreira, J.~Christiansen, and K.~Wyllie, ``{GBTx Manual - GBT project},''
  \emph{Manual v0.16 Draft}, pp. 15--16, 2018.

\bibitem{aliceo2readoutcard}
P.~Boeschoten, K.~Alexopoulos, B.~von Haller, A.~Wegrzynek, D.~Berzano
  \emph{et~al.}, ``{ReadoutCard},''
  \url{https://github.com/AliceO2Group/ReadoutCard}, tag v0.22.2, commit
  c8301df.

\bibitem{aliceo2LLA}
K.~Alexopoulos, ``{LLA (Low-Level Arbitration)},''
  \url{https://github.com/AliceO2Group/LLA}, tag v0.1.0, commit ec7b6d3.

\bibitem{directcuserguide}
Microsemi, ``{DirectC v4.1 User Guide},'' 2019.

\bibitem{stapl1999}
{JEDEC Solid State Technology Association}, ``{Standard Test and Programming
  Language (STAPL)},'' 1999.

\bibitem{actelstapl2003}
Actel, ``{Application Note AC171: ISP and STAPL},'' 2003.

\bibitem{actelstapl2008}
{Actel STAPL Player v1.1 Release Notes (2008)}.
  \url{http://soc.microsemi.com/download/program$\_$debug/stapl/stapl11.aspx}.

\bibitem{lhcb}
F.~Alessio and J.~V.~V. Barbosa, LHCb team, private communication.

\bibitem{gbtscamanual}
S.~Bonacini, A.~Caratelli, R.~Francisco, K.~Kloukinas, A.~Marchioro,
  P.~Moreira, and C.~Paillard, ``{GBT-SCA, The Slow Control Adapter ASIC for
  the GBT System, User Manual v8.2 - Draft},'' pp. 35--40, April 2017.

\bibitem{sol402018barbosa}
J.~V.~V. Barbosa, C.~Gaspar, and F.~Alessio, ``{The new version of the LHCb
  SOL40-SCA core to drive front-end GBT-SCAs for the LHCb upgrade},''
  \emph{PoS}, p. 078, 2018.

\end{thebibliography}
	\end{flushleft}
	
\end{document}